\documentclass[10pt, conference]{IEEEtran}

\usepackage{booktabs} 
\usepackage{cite}

\usepackage{setspace,multirow}
\usepackage{algpseudocode}
\usepackage{mathrsfs}
\usepackage{amsmath,amsfonts,amssymb,amsthm}
\usepackage{algorithm}
\usepackage{graphicx}
\usepackage{array}
\usepackage{caption}
\usepackage{makecell}

\usepackage[caption=false,font=footnotesize]{subfig}

\usepackage{epstopdf}
\usepackage[hyphens]{url}
\usepackage[dvipsnames]{xcolor}
\usepackage[normalem]{ulem}

\newtheorem{remark}{Remark}

\algrenewcommand\algorithmicrequire{\textbf{Input:}}
\algrenewcommand\algorithmicensure{\textbf{Output:}}

\newcommand{\bX}{\mathbf{X}}

\usepackage{glossaries} 

\newcommand{\bx}{\mathbf{x}}

\newcommand{\balpha}{\boldsymbol{\alpha}}

\usepackage{geometry}
\usepackage[T1]{fontenc}
\pdfoutput=1
\IEEEoverridecommandlockouts
\makeatletter
\def\endthebibliography{%
	\def\@noitemerr{\@latex@warning{Empty `thebibliography' environment}}%
	\endlist
}
\makeatother

\usepackage{glossaries}

\usepackage{tikz}
\usetikzlibrary{shapes.arrows}
\usetikzlibrary{plotmarks}
\usetikzlibrary{patterns}
\usetikzlibrary{arrows.meta}
\usetikzlibrary{positioning}
\usetikzlibrary{spy}
\usepackage{pgfplots}
\usepgfplotslibrary{colorbrewer}
\usepgfplotslibrary{fillbetween}

\IEEEoverridecommandlockouts
\glsenableentrycount
\makeglossaries

\newacronym{gsl}{GSL}{ground to satellite link}
\newacronym{gsd}{GSD}{ground sample distance}
\newacronym{fov}{FoV}{field of view}
\newacronym{gtfp}{GTFP}{ground track frame period}
\newacronym{gs}{GS}{ground station}
\newacronym{snr}{SNR}{signal-to-noise ratio}

\newcommand{\phuc}[1]{\textcolor{black}{#1}}

\begin{document}
	
	\title{{Scalable Data Transmission Framework for Earth Observation Satellites with Channel Adaptation}}

\author{Van-Phuc Bui, Shashi Raj Pandey, Israel Leyva-Mayorga, and Petar Popovski\\ \IEEEauthorblockA{Department of Electronic Systems, Aalborg University, Denmark (\{vpb,  srp, ilm, petarp\}@es.aau.dk)\\
 }\thanks{This work was supported by the Villum Investigator Grant ``WATER'' from the Velux Foundation, Denmark.}}

	\maketitle
\begin{abstract}
The immense volume of data generated by Earth observation (EO) satellites presents significant challenges in transmitting it to Earth over rate-limited satellite-to-ground communication links. This paper presents an efficient downlink framework for multi-spectral satellite images, leveraging adaptive transmission techniques based on pixel importance and link capacity. By integrating semantic communication principles, the framework prioritizes critical information, such as changed multi-spectral pixels, to optimize data transmission. The process involves preprocessing, assessing pixel importance to encode only significant changes, and dynamically adjusting transmissions to match channel conditions. Experimental results on the real dataset and simulated link demonstrate that the proposed approach ensures high-quality data delivery while significantly reducing number of transmitted data, making it highly suitable for satellite-based EO applications.
\end{abstract}

\begin{IEEEkeywords}
Satellite communication, Change detection, Image processing.
\end{IEEEkeywords}

\section{Introduction} \label{sec:Intro}
Earth observation (EO) through remote sensing satellites play an indispensable role in monitoring environmental conditions by offering rapid and wide-ranging coverage of specific areas. This capability enables diverse applications, including land use mapping, urban development analysis, and disaster response through the acquisition of imagery \cite{poursanidis2017remote, nguyen2024emerging}. However, the deployment of high-resolution sensors generates vast amounts of data, which demand substantial communication bandwidth and onboard storage for effective delivery to ground stations. The traditional method of transmitting raw images to Earth for processing and distribution has become inefficient due to the exponential growth in data volumes from modern EO missions. For example, the Sentinel-2 satellite system produces a staggering 2.4 Terabits of data daily, capturing global surface images at five-day intervals \cite{sentinel2}. While delivering the entire captured data allows for comprehensive analysis, it poses challenges related to storage and transmission capacities. Consequently, new satellite missions are integrating onboard data processing to transfer data processing tasks from ground-based systems to satellites. These advancements aim to enhance transmission efficiency, reduce memory usage and communication costs, and optimize the selection of relevant data for downlink \cite{trautner2010ongoing}.

With recent advancements in remote sensing and the growing application of artificial intelligence in space information networks, the role of machine learning in optimizing these systems is becoming increasingly prominent. Deep learning techniques, in particular, have demonstrated exceptional performance in image processing tasks, which has motivated researchers to unleash its full potential for EO tasks. For semantic image segmentation, Fully Convolutional Neural Networks (FCNNs), especially those with encoder-decoder architectures like U-Net, have shown great promise for automatically generating ground truths \cite{10659182}. The reason is that these networks excel at pixel-wise classification, effectively mapping each pixel in an image to a specific semantic category, enabling the creation of accurate and detailed annotations with minimal human effort.
In the domain of change detection, an area that dates back to the early stages of aerial imaging \cite{hussain2013change}, Convolutional Neural Networks (CNNs) have been widely adopted for analyzing and comparing images across various scenarios \cite{zagoruyko2015learning}. 
Resource allocation in Space Information Systems has also been explored. For example, \cite{WangShengZhuangJSAC2018} and \cite{WangShengYeArvix2019} address scheduling problems to maximize task priority or network utility. Similarly, \cite{tran2022satellite, nguyen2023security} investigated joint resource allocation and cache placement or security, highlighting the importance of resource coordination. These works, however, assume that all collected satellite data must be transmitted, which may not always be feasible in real-world scenarios. 
In a recent work~\cite{10659182}, the authors developed a semantic data encoding system based on change detection, ensuring that up to $95\%$ of critical information is transmitted effectively. However, a limitation of this approach is its inability to adapt the transmission rate according to the varying channel capacities between the satellite and the gateway, which frequently fluctuate. This paper builds upon these advancements by proposing a novel framework for efficient downlink communication of satellite images. Specifically, we propose a deep learning-based algorithm to predict channel capacity for each connection and dynamically prioritize the selection of the most critical pixels for transmission. This ensures that the amount of data to transmit aligns with the available link capabilities, enabling satellites to adaptively adjust to fluctuating link conditions. Our contributions are summarized as follows.
\begin{itemize}
    \item We propose a novel deep learning-based semantic encoding framework that prioritizes critical information from multi-spectral satellite images based on varying channel conditions,  thereby reducing unnecessary data overhead.
    \item Our approach integrates channel capacity prediction to dynamically adjust the transmission of encoded data, ensuring efficient use of satellite-to-gateway communication links under fluctuating conditions.
    \item Experiments on real dataset demonstrate the effectiveness of the proposed method in maintaining high-quality data transmission while significantly optimizing energy consumption and bandwidth utilization. This bridges the gap between semantic encoding and real-time channel adaptation, providing a robust framework for next-generation satellite communication systems.
\end{itemize}

\section{System Model}

\begin{figure}[!t]
    \centering
    \includegraphics[width=0.5\textwidth]{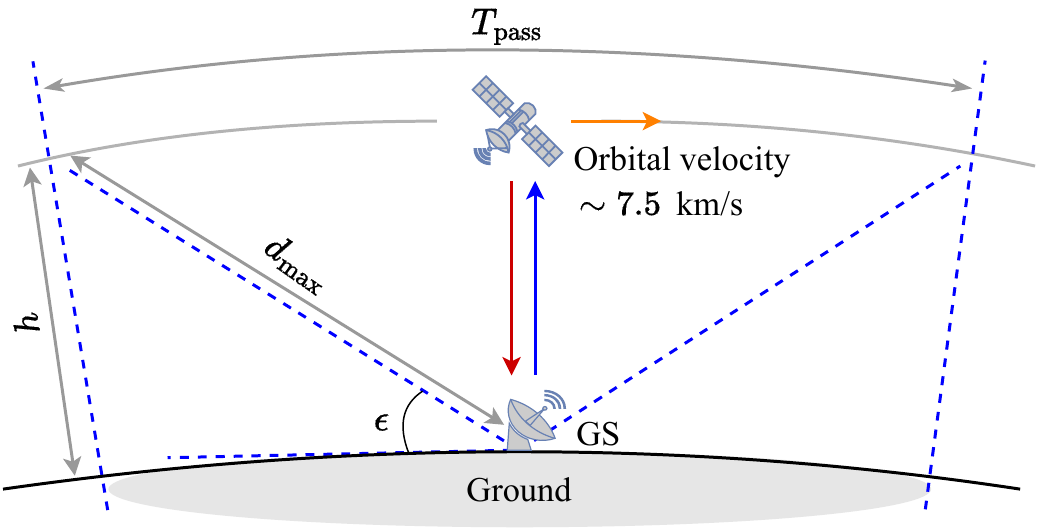}
    \caption{The system model.}
    \label{fig_system}
    \vspace{-0.5cm}
\end{figure}

We consider a satellite communication framework depicted in Fig.~\ref{fig_system}, where a Low Earth Orbit (LEO) satellite communicates with a ground station (GS) directly connected to a server. The satellite captures multi-spectral images and utilizes an onboard artificial intelligence (AI) module for feature extraction. 
We examine two coregistered multi-spectral satellite images captured at different time intervals: $ \bX^{t_0} = \{x^{t_0}(i,j,k) \mid 1 \leq i \leq H,\ 1 \leq j \leq W,\ 1 \leq k \leq D\} $ and $ \bX^{t_1} = \{x^{t_1}(i,j,k) \mid 1 \leq i \leq H,\ 1 \leq j \leq W,\ 1 \leq k \leq D\} $, where $ H $, $ W $, and $ D $ represent the height, width, and number of spectral bands of the images, respectively. Here, $ \bX^{t_0} $ serves as the reference image, and $ \bX^{t_1} $ is the newly acquired image. Our objective is to transmit only the changed multi-spectral pixels (MPs) from $ \bX^{t_1} $ to the ground station, while adhering to the constraints of channel capacity. Let $ \mathcal{S} = \{s_{ij}\} $ denote the change map, where $ s_{ij} \in \{0,1\} $ indicates whether pixel $ (i,j) $ has changed ($ s_{ij} = 1 $) or remains unchanged ($ s_{ij} = 0 $). A pixel change is defined as a temporal variation at a specific location in the coregistered images, encompassing events such as land-use changes, urban development, deforestation, or other comparable changes \cite{8418840}.

We denote the set of pixels identified as changed for transmission as $ \mathcal{P} \subseteq \mathcal{X}^{t_1} $, where $ \mathcal{X}^{t_1} = \{\bx_{ij}^{t_1}\} $ represents all pixels in $ \bX^{t_1} $. The selection of $ \mathcal{P} $ must take into account the communication window capacity constraints and prioritize the transmission of critical information.

\subsection{Communication and Capacity Constraints}
The satellite communicates with the gateway through a downlink channel characterized by additive-white Gaussian noise (AWGN) and free-space path loss. In this paper, we consider the simplest case of periodic image arrivals with no buffering, where the transmission strategy ensures that each image is immediately sent within the satellite's visibility duration. In this scenario, the periodic image arrival rate is aligned with the transmission rate throughout the next satellite pass
to maintain a consistent minimal quality for all images. Since no buffering is assumed, images are transmitted sequentially without delay, ensuring that the system operates in a predictable and stable manner. This eliminates the need for complex scheduling or queuing mechanisms, making it well-suited for scenarios with strict timing requirements. The approach ensures efficient use of the available transmission window while adhering to periodic arrival constraints.
The time available for data transmission is determined by the satellite's visibility duration with respect to the GS.
\subsubsection{Visibility Duration}
The visibility duration, $T^\text{pass}$, represents the maximum time during which a satellite remains visible to a ground station and can transmit data. This duration is determined by the satellite's orbital altitude $h$, the Earth's radius $R_\text{Earth}$, and the minimum elevation angle $\epsilon$, which defines the lowest angle above the horizon at which the satellite can be observed. The inclusion of $\epsilon$ ensures that line-of-sight communication is not obstructed by the Earth's curvature and other obstacles. 
First, the maximum slant range distance $d_\text{max}$ between satellite and GS can be computed by 
\begin{IEEEeqnarray}{lll}\label{dmax}
d_\text{max} = \sqrt{(R_\text{Earth} + h)^2 - (R_\text{Earth} \cos \epsilon)^2} - R_\text{Earth} \sin \epsilon,
\end{IEEEeqnarray}
which can be obtained by using cosines law for triangle of satellite, GS, and the Earth's center. We observe that \eqref{dmax} highlights the geometric dependency of the visibility duration on the minimum elevation angle $\epsilon$. A higher $\epsilon$ reduces atmospheric effects, such as signal attenuation and noise, by avoiding low-angle transmissions where the signal path through the atmosphere is longest. Conversely, a lower $\epsilon$ maximizes $T^\text{pass}$ and allows more time for data transmission, but at the cost of increased atmospheric interference. 
Once $d_\text{max}$ is determined, the visibility duration $T_\text{pass}$ can be calculated using the following Remark.
\begin{remark}
The visibility duration $T_\text{pass}$ is given by:
\begin{IEEEeqnarray}{lll}\label{Tpass}
T_\text{pass} = \frac{2 \left( R_\text{Earth} + h \right)}{v_\text{orb}} \arcsin \left( \frac{d_\text{max} \cos \epsilon}{R_\text{Earth} + h} \right),
\end{IEEEeqnarray}
where $v_\text{orb}$ represents the satellite's orbital velocity.
\end{remark}
\begin{proof}
Let $\beta$ denote half of the central angle subtended by the satellite’s trajectory that remains within visibility. The visibility duration $T_\text{pass}$ can then be expressed as
\begin{IEEEeqnarray}{lll}\label{lemma_proof_1}
T_\text{pass} = \frac{2 \left( R_\text{Earth} + h \right)}{v_\text{orb}} 2\beta.
\end{IEEEeqnarray}
To derive $\beta$, we use the geometric relationship between the elevation angle, the slant range, and the central angle, which is given by $d_\text{max}\cos\epsilon = (R_\text{Earth}+h)\sin\beta$ \cite{gordon1993principles}. Substituting $\beta$ into \eqref{lemma_proof_1} yields \eqref{Tpass}, thus completing the proof.
\end{proof}
We note that the duration $T^\text{pass}$ determines the window of opportunity for data transmission, making it a critical parameter in satellite communication system design.  

\begin{figure}[!t]
    \centering
    \input{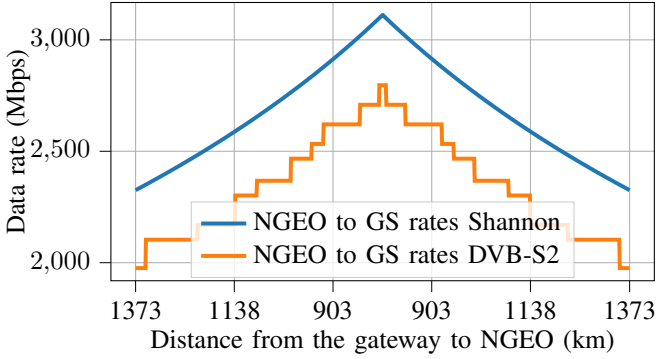}
    \caption{Achievable data rates for satellite-to-ground communication in one orbit.}
    \label{fig:rates_comparison}
    \vspace{-10pt}
\end{figure}
\subsubsection{Data Rate and Link Capacity}
During  $T^\text{pass}$, the achievable data rate depends on the signal-to-noise ratio (SNR) at time $t$, denoted as $\gamma_t$. The SNR for the satellite-to-ground link is expressed as
\begin{IEEEeqnarray}{lll}
\label{eq_SNR}
    \gamma_t = G_{\text{tx}}G_{\text{rx}}P_{\text{tx}}\left(\frac{c}{4\pi d_t f_c \varpi}\right)^2,
\end{IEEEeqnarray}
where $G_{\text{tx}}$ and $G_{\text{rx}}$ are the transmitter and receiver antenna gains, $P_{\text{tx}}$ is the satellite transmission power, $f_c$ is the carrier frequency, $c$ is the light speed, $d_t$ is the satellite-to-gateway distance at time $t$, and $\varpi^2$ is the noise power. The distance $d_t$ changes dynamically as the satellite orbits Earth. Once $\gamma_t$ is known, the data rate $R_t$ at time $t$ is calculated based on the SNR and the selected modulation and coding scheme (MCS). Using the DVB-S2 system \cite{morello2006dvb}, the achievable rates $\bar{R}$ and their corresponding minimum SNR requirements $\gamma_\text{min}(\bar{R})$ are defined in a set
\begin{IEEEeqnarray}{lll}
    \mathcal{Q}_\text{DVB-S2} = \{(\bar{R}, \gamma_\text{min}(\bar{R}))\}.
\end{IEEEeqnarray}
The rate $R_t$ for a given time $t$ is determined as
\begin{IEEEeqnarray}{lll}
    R_t = \max \big\{\bar{R} \in \mathcal{Q}_\text{DVB-S2} : \gamma_t \geq \gamma_\text{min}(\bar{R})\big\}.
\end{IEEEeqnarray} 
The total capacity for one orbital pass, $C_\text{orbit}$, is calculated as $C_\text{orbit} = \int_0^{T^\text{pass}} R_t \, dt$.
To approximate $C_\text{orbit}$, the horizon $T^\text{pass}$ is discretized into $N$ intervals, where each interval $k$ is assumed as a static channel, allowing the instantaneous rate $R_t$ to be approximated by a constant rate $R_k$. This yields the discrete form 
\begin{IEEEeqnarray}{l}
    C_\text{orbit} = \sum_{k=1}^N R_k \Delta t_k,\label{eq_Corbit}
\end{IEEEeqnarray}
where $\Delta t_k$ is the duration of interval $k$ and $\sum_{k=1}^N \Delta t_k = T^\text{pass}$. 
Fig.~\ref{fig:rates_comparison} shows the variation in achievable data rates during a single orbital pass. This figure illustrates the theoretical Shannon rates and the practical rates achievable using the DVB-S2 system as functions of time during a single orbital pass.

\subsection{Problem Formulation}
To address the need for prioritizing critical information in satellite communication,  we define a utility function $f(\mathbf{X}^{t_1}, \hat{\mathbf{X}}^{t_1})$ based on Peak Signal-to-Noise Ratio (PSNR), which measures the similarity between the original transmitted image $\mathbf{X}^{t_1}$ and the reconstructed image $\hat{\mathbf{X}}^{t_1}$. The optimization problem is formulated as
\begin{IEEEeqnarray}{rlll} \label{origin_prob}
	\text{P1}: \max_{\balpha} &~~ f(\mathbf{X}^{t_1}, \hat{\mathbf{X}}^{t_1}),  \IEEEyesnumber\IEEEyessubnumber\label{origin_prob_a} \\
    & ~~\sum_{(i,j)} \alpha_{ij} b_{ij} \leq C_\text{orbit}, \IEEEyessubnumber\label{origin_prob_b},
\end{IEEEeqnarray}
where $\alpha_{ij}$ is a binary decision variable where $\alpha_{ij} = 1$ if MP $\mathbf{x}_{ij}^{t_1}$ is selected for transmission, and $\alpha_{ij} = 0$ otherwise. $b_{ij}$ corresponds to the number of bits required to encode $\mathbf{x}_{ij}^{t_1}$. \phuc{While \eqref{origin_prob} represents a conceptually ideal formulation, it is practically infeasible in satellite scenarios to measure $f(\mathbf{X}^{t_1}, \hat{\mathbf{X}}^{t_1})$ instantaneously, as PSNR is typically evaluated post-hoc rather than during real-time operations. From the PSNR formula, maximizing PSNR involves prioritizing the transmission of pixels with the greatest changes in value, as these dominate the reconstruction error. To approximate this, we introduce an importance score $p_{ij}$, which quantifies the magnitude of change for each MP $\bx_{i,j}$ and serves as a practical proxy for the pixel-level reconstruction error. With this, \eqref{origin_prob} is reformulated to focus on maximizing the cumulative importance score of the transmitted pixels while adhering to the channel capacity constraint, which is mathematically expressed as
\begin{IEEEeqnarray}{rlll} \label{new_problem_priority}
	\text{P2}: \max_{\balpha} & ~~\sum_{(i,j)} p_{ij} \alpha_{ij} \IEEEyesnumber\IEEEyessubnumber\label{new_problem_priority_a} \\
    \mathrm{s.t.} & ~~\alpha_{ij} \leq s_{ij}, \IEEEyessubnumber\label{new_problem_priority_b}\\
    & ~~\sum_{(i,j)} \alpha_{ij} b_{ij} \leq C_\text{orbit} \IEEEyessubnumber\label{new_problem_priority_c}.
\end{IEEEeqnarray}}
The constraint \eqref{new_problem_priority_b} ensures that only changed pixels, as indicated by the change map $s_{ij}$, are considered for transmission, where $s_{ij} = 1$ denotes a change and $s_{ij} = 0$ otherwise. This optimization problem ensures that high-priority pixels with large $p_{ij}$  values are selected, while adhering to the channel capacity constraint, ensuring efficient and adaptive satellite communication.

\section{ML-based Data Encoding}

As observed in \eqref{eq_SNR}, the satellite SNR can be accurately predicted in advance by analyzing the variations in the distance between the ground station GS and the satellite during each orbit. Consequently, this allows for the estimation of the orbit capacity, as described in \eqref{eq_Corbit}. This predictability is of significant importance as it serves as a foundation for determining the optimal amount of data to be transmitted to the ground station. To address the challenge of efficiently transmitting critical information from satellite-captured multi-spectral images \eqref{new_problem_priority}, we propose a three-stage method that includes preprocessing, building a Change scoring map by using Change-Net, and adaptively selecting the most significant pixels for transmission based on the predicted $C_{\text{orbit}}$. 
\subsubsection{Preprocessing}
The preprocessing step aims to prepare multi-spectral image data for effective change detection by addressing noise, illumination inconsistencies, and cloud cover. First, a subset of $N$ spectral bands $(N < D)$ is selected based on their relevance to the detection of changes, such as visible and near-infrared bands, to reduce computational requirements. Subsequently, radiometric normalization is applied to mitigate variations caused by differences in imaging conditions, including sun angle, light intensity, and atmospheric effects. The employed method utilizes relative radiometric normalization, based on the z-score technique, which standardizes images by adjusting their mean to zero and their standard deviation to one. For a given pair of MP images, represented as $\bx_{ij}^{t_0} = [x_1^{t_0}, x_2^{t_0}, \dots, x_N^{t_0}]$ and $\bx_{ij}^{t_1} = [x_1^{t_1}, x_2^{t_1}, \dots, x_N^{t_1}]$, the normalization process is described as $\hat{x}_n^{t_0} = {x_n^{t_0} - \mu_{\bx_n^{t_0}}}/{\sigma_{\bx_n^{t_0}}^2}$ and $\hat{x}_n^{t_1} = {x_n^{t_1} - \mu_{\bx_n^{t_1}}}/{\sigma_{\bx_n^{t_1}}^2}$, where $\mu_{\bx_n^{t_k}}$ and $\sigma_{\bx_n^{t_k}}^2$ represent the mean and variance, respectively, for band $n$ in image $\mathbf{X}^{t_k}$, with $k \in \{0, 1\}$. This normalization procedure effectively mitigates radiometric discrepancies between the multi-spectral images that arise due to varying imaging conditions.

\subsubsection{Change Scoring Map Construction}
Given two multi-spectral images that are geometrically registered, uniformly illuminated, our objective is to develop a computationally efficient automatic change detection method suitable for satellite applications. Inspired by the works of \cite{10659182, ronneberger2015u}, we design Change-Net, an architecture based on U-Net, to execute semantic segmentation on images. A negative log-likelihood loss function is employed to classify each measurement point MP in the observed images as changed or unchanged. Specifically, we consider a change detection training set comprising tuples $\mathcal{D}_\text{change}$ defined as
\begin{IEEEeqnarray}{lll}
\mathcal{D}_\text{change} = \left\{\,(\mathbf{b}_{i} , y_{i}) \in \mathcal{B}\times \mathcal{Y} \,\Big|\, i = 1, \dots, M_\text{change} \,\right\},
\end{IEEEeqnarray}
where $\mathbf{b}_{i}$ represents the inputs and $y_{i}$ the corresponding labels. Denoting $\boldsymbol{\theta}$ as the set of learnable parameters to be optimized, the likelihood function $p(\mathbf{b}_{i},\boldsymbol{\theta})$ is the joint probability density of the observed data, which acts as a mapping from the inputs $\mathbf{b}_{i}$ to outputs $y_{i}$. Letting $m$ denote the mini-batch size, the optimal parameter set $\boldsymbol{\theta}_\text{model}$ is obtained during training as
\begin{IEEEeqnarray}{lll}
\boldsymbol{\theta}_\text{model} = \underset{\boldsymbol{\theta}}{\arg\max}\prod_{i=1 }^m p(y_i|\mathbf{b}_{i},\boldsymbol{\theta}).
\end{IEEEeqnarray}

In our context, the system output $\Bar{s}_{ij} \in [0,1]$ represents the change score for the cloud-removed measurement point $\tilde{\mathbf{x}}^{t_1}_{ij}$, calculated as the probability
\begin{equation}
p(s_{ij}=1|\tilde{\mathbf{x}}^{t_1}_{ij},\boldsymbol{\theta}_\text{model}) = \xi\left(\boldsymbol{\theta}_\text{model}^\top \tilde{\mathbf{x}}^{t_1}_{ij}\right),
\end{equation}
where $\xi(\cdot)$ denotes the Log Softmax function. A value of $\Bar{s}_{ij}$ close to 1 indicates a high likelihood of change at location $(i,j)$. Therefore, to achieve accurate predictions, we aim to develop a scoring system defined by
\begin{IEEEeqnarray}{rll}\label{score_update}
s_{ij}^p = \mathbf{1}(\Bar{s}_{ij} \geq \tau),
\end{IEEEeqnarray}
where $\tau$ is a predefined threshold, experimentally determined through an iterative process using captured images and orbit capacity to satisfy constraint \eqref{new_problem_priority_c}. We utilize the FC-EF-Res structure \cite{daudt2019multitask} within the encoder-decoder block for training on the dataset. Specifically, a pair of 4-band images (RGB and Near-Infrared bands) is used as input to Change-Net. Unlike conventional classification models that attempt to label each measurement point in an image as changed or unchanged, our Change-Net produces a change probability map. This map is then processed in a scoring phase to generate a semantic map containing the compacted and transmitted measurement points. This procedural approach is designed to mitigate the inefficiencies present in existing change detection methods.

\subsubsection{Adaptive Pixel Selection Based on Channel Capacity}
Given the predicted channel capacity $C_{\text{orbit}}$ during the satellite's visibility window, the pixel selection process prioritizes those with the highest change scores, ensuring that the data transmitted fits within the capacity constraint. MPs are ranked by their scores $s_{ij}^p,$ and the top-ranked MPs are selected such that \eqref{new_problem_priority_c} is satisfied. 

\section{Numerical Results}
\begin{table}[!t]
\caption{Simulation Parameters}
\resizebox{8.5cm}{!} 
{\begin{tabular}{ll}
		\hline
		Parameter & Value \\
		\hline
        Altitude ($h$) & 600 km \\
        Min. elevation angle ($\epsilon$) & 30° \\
        Speed of light ($c$) & $2.997 \times 10^8$ m/s \\
        Gravity constant ($G$) & $6.673 \times 10^{-11}$ m$^3$/kg$\cdot$s$^2$ \\
        Earth's mass ($M_e$) & $5.9736 \times 10^{24}$ kg \\
        Boltzmann constant ($k$) & $1.38 \times 10^{-23}$ J/K \\
        Satellite antenna gain $(G_\text{tx})$ & 32.13 dBi \\
        Gateway antenna gain $(G_\text{rx})$ & 34.2 dBi \\
        Antenna efficiency ($\eta$) & 0.55 \\
        Noise temp. ($T_n$) & 290 K \\
        Carrier freq. ($f$) & 26 GHz \\
        Bandwidth ($B$) & 500 MHz \\
        Tx/Rx antenna diam. ($A_\text{rx}, A_\text{tx}$) & (0.26 m, 0.26 m) \\
        Transmission power $(P_\text{tx})$ & 10 W      \\
		\hline
	\end{tabular}
}
\label{parameters}
\end{table}

To evaluate the proposed algorithm, we use the Onera Satellite Change Detection (OSCD) data set \cite{dataset_change_detection}, publicly available on IEEE DataPort, and widely used for detecting changes in Sentinel-2 multi-spectral images \cite{daudt2018urban}. The data set provides multis-pectral imagery, facilitating model training by mitigating cold-start issues. Classical preprocessing is applied, and class imbalance is addressed by assigning weights inversely proportional to class frequencies. The data set is divided into $60\%$ training data and $40\%$ testing data using shuffled and stratified sampling. Experiments are conducted on a 28-core Intel Xeon Gold @ 2.8~GHz server with 256~GB memory and an NVIDIA Tesla V100 GPU, using PyTorch for implementation. Additional parameters are listed in Table~\ref{parameters}.

\begin{figure*}[thp]

\begin{minipage}{0.12\textwidth}
\includegraphics[trim=0.cm 0.0cm 0.cm 0cm, clip=true, width=0.77in]{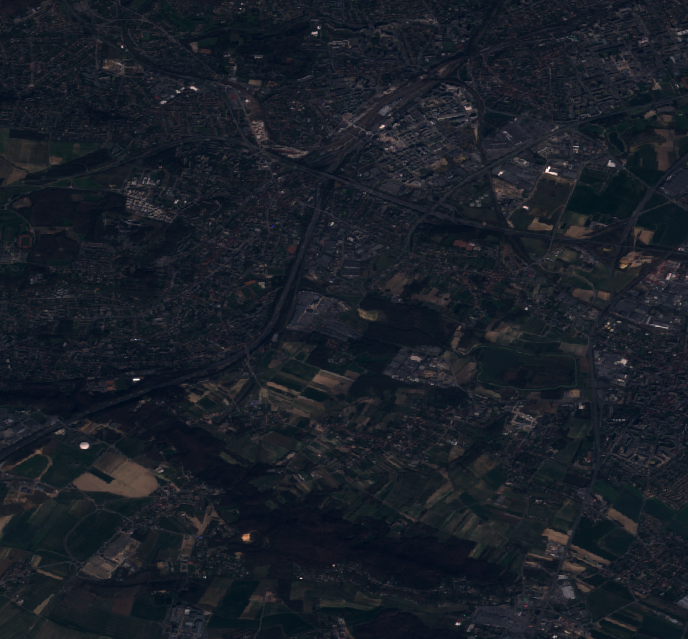} \\ 
\vspace*{-0pt}
\end{minipage}
\begin{minipage}{0.12\textwidth}
\includegraphics[trim=0cm 0cm 0cm 0cm, clip=true, width=0.8in]{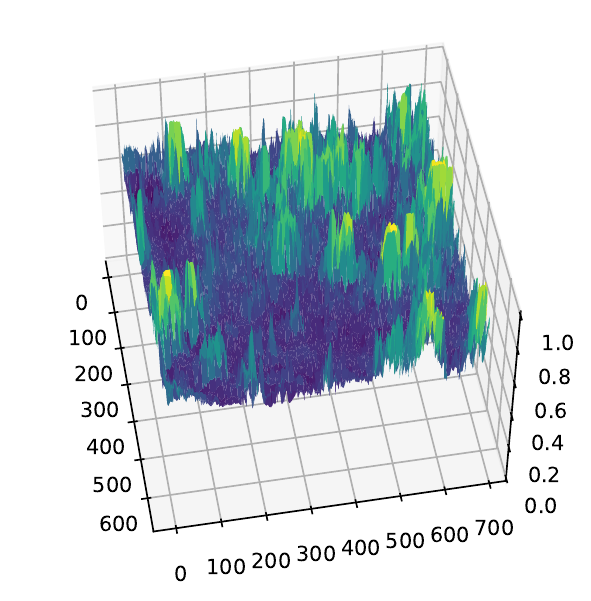} \\ 
\vspace*{-5pt}
\end{minipage}
\begin{minipage}{0.12\textwidth}
\includegraphics[trim=0.cm 0.0cm 0.cm 0cm, clip=true, width=0.77in]{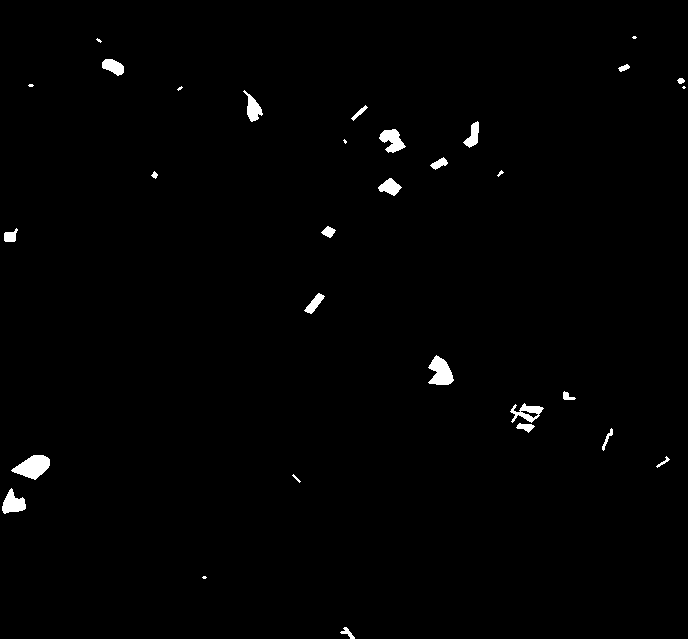} \\ 
\vspace*{-0pt}
\end{minipage}
\begin{minipage}{0.12\textwidth}
\includegraphics[trim=0.cm 0.0cm 0.cm 0cm, clip=true, width=0.8in]{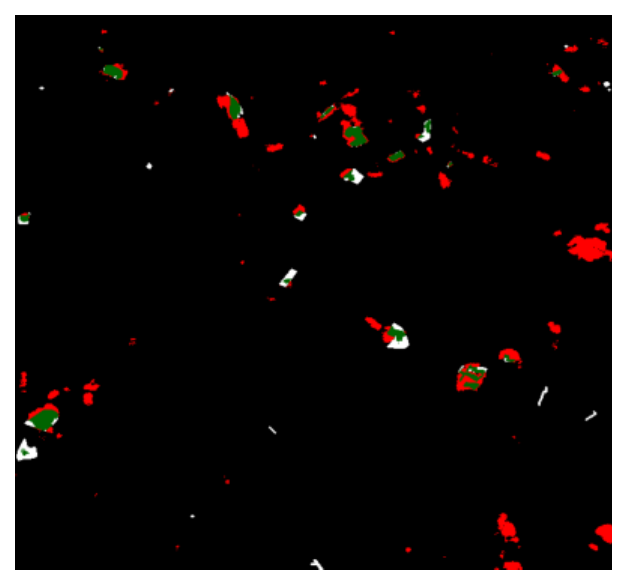} \\ 
\vspace*{-0pt}
\end{minipage}
\begin{minipage}{0.12\textwidth}
\includegraphics[trim=0.cm 0.0cm 0.cm 0.0cm, clip=true, width=.8in]{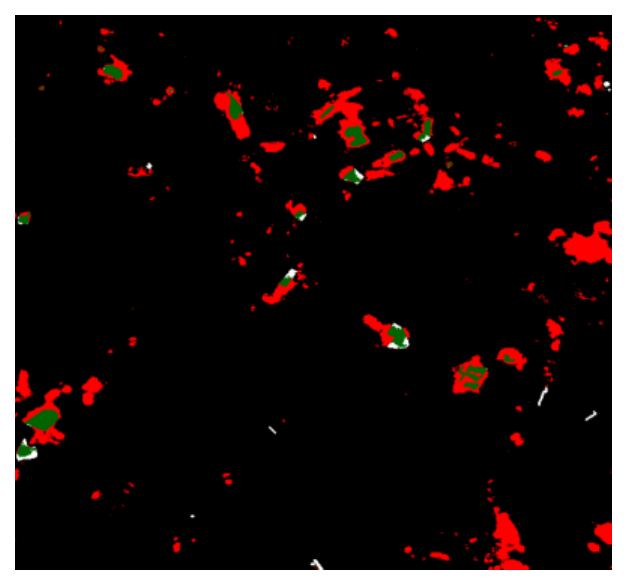} \\ 
\vspace*{-0pt}
\end{minipage}
\begin{minipage}{0.12\textwidth}
\includegraphics[trim=0.cm 0.0cm 0.cm 0.0cm, clip=true, width=.8in]{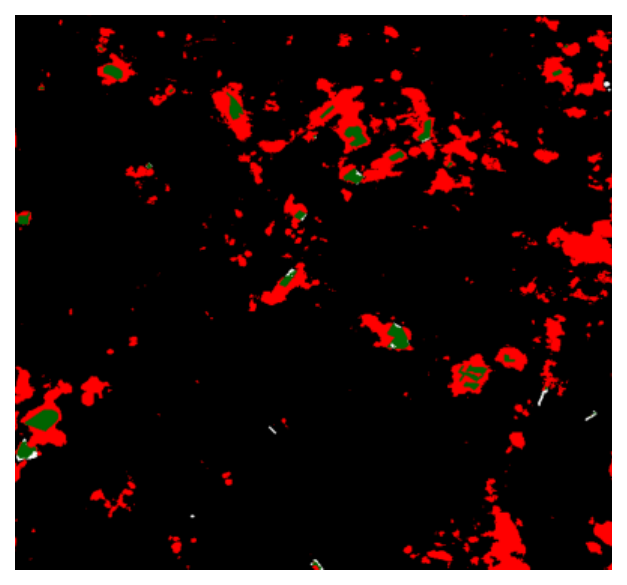} \\ 
\vspace*{-0pt}
\end{minipage}
\begin{minipage}{0.12\textwidth}
\includegraphics[trim=0.cm 0.0cm 0.cm 0.0cm, clip=true, width=.8in]{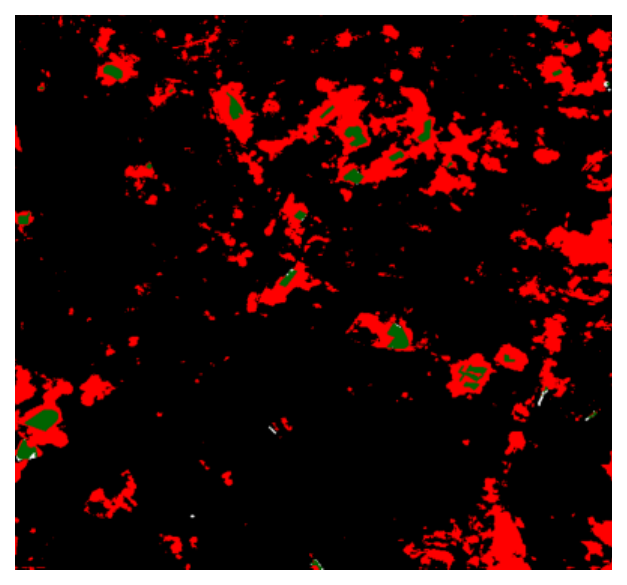} \\ 
\vspace*{-0pt}
\end{minipage}
\begin{minipage}{0.12\textwidth}
\includegraphics[trim=0.cm 0.0cm 0.cm 0.0cm, clip=true, width=.8in]{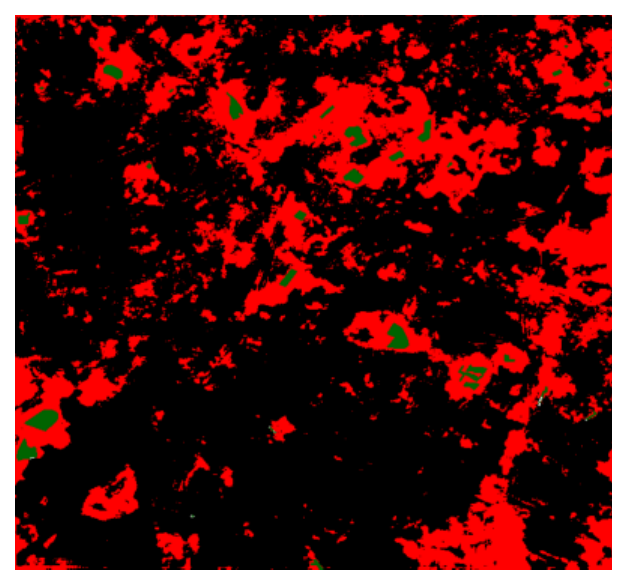} \\ 
\vspace*{-0pt}
\end{minipage}

\begin{minipage}{0.12\textwidth}
\includegraphics[trim=0.cm 0.0cm 0.cm 0cm, clip=true, width=0.77in]{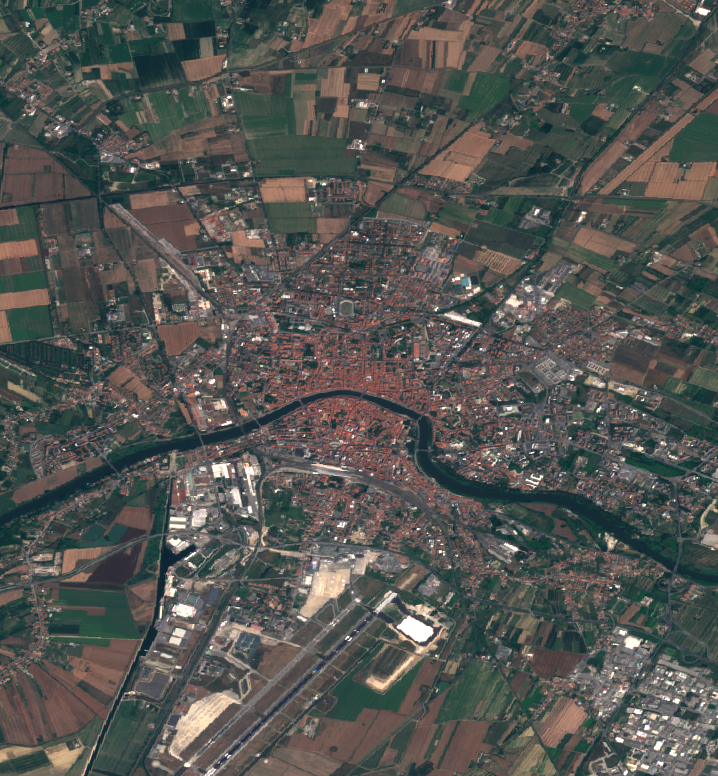} \\ 
\vspace*{-0pt}
\end{minipage}
\begin{minipage}{0.12\textwidth}
\includegraphics[trim=0cm 0cm 0cm 0cm, clip=true, width=0.8in]{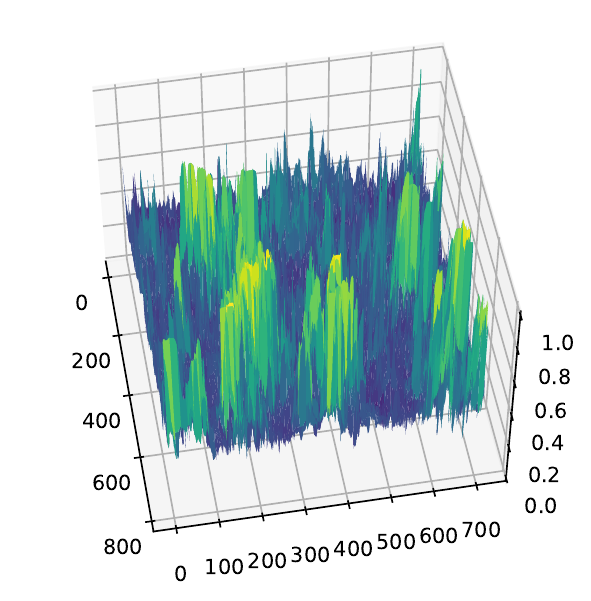} \\ 
\vspace*{-5pt}
\end{minipage}
\begin{minipage}{0.12\textwidth}
\includegraphics[trim=0.cm 0.0cm 0.cm 0cm, clip=true, width=0.77in]{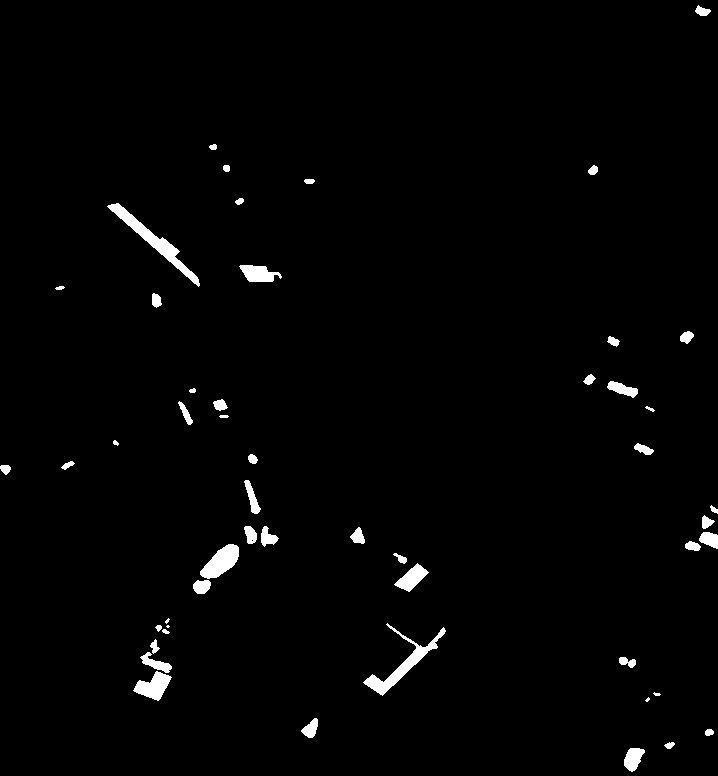} \\ 
\vspace*{-0pt}
\end{minipage}
\begin{minipage}{0.12\textwidth}
\includegraphics[trim=0.cm 0.0cm 0.cm 0cm, clip=true, width=0.8in]{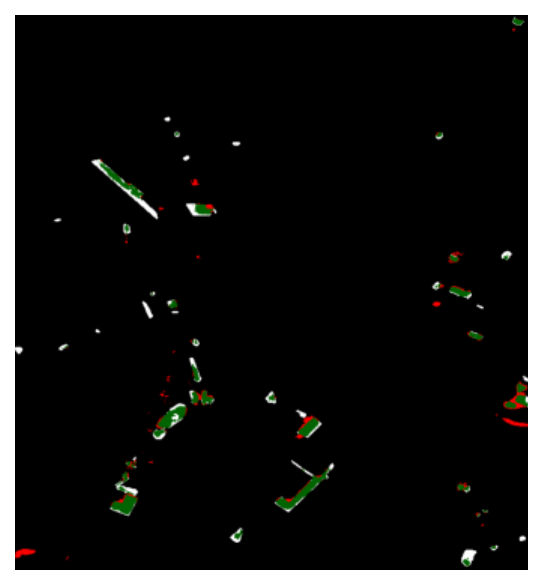} \\ 
\vspace*{-0pt}
\end{minipage}
\begin{minipage}{0.12\textwidth}
\includegraphics[trim=0.cm 0.0cm 0.cm 0.0cm, clip=true, width=.8in]{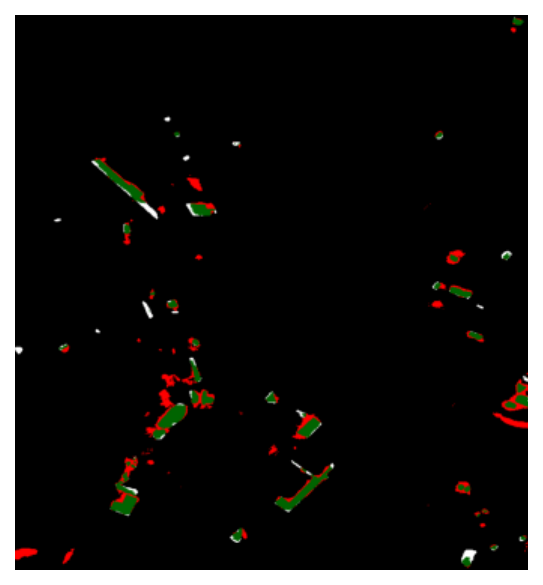} \\ 
\vspace*{-0pt}
\end{minipage}
\begin{minipage}{0.12\textwidth}
\includegraphics[trim=0.cm 0.0cm 0.cm 0.0cm, clip=true, width=.8in]{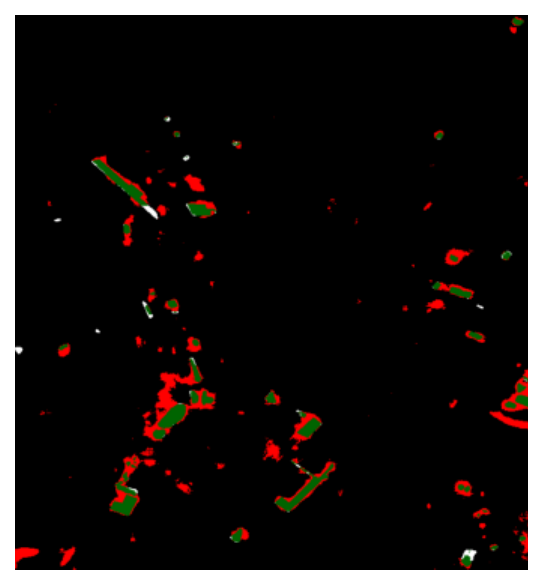} \\ 
\vspace*{-0pt}
\end{minipage}
\begin{minipage}{0.12\textwidth}
\includegraphics[trim=0.cm 0.0cm 0.cm 0.0cm, clip=true, width=.8in]{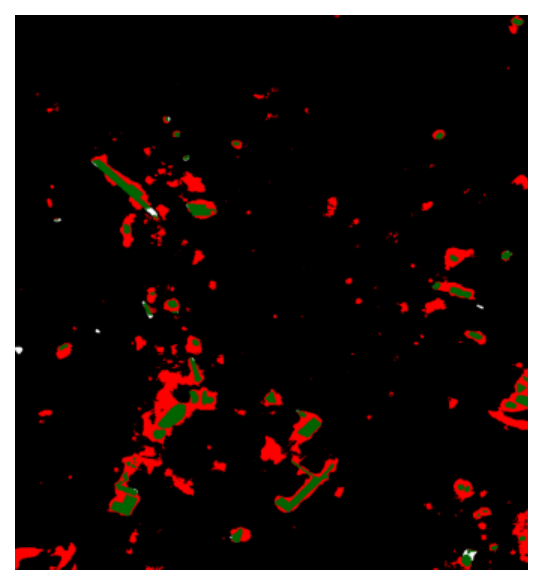} \\ 
\vspace*{-0pt}
\end{minipage}
\begin{minipage}{0.12\textwidth}
\includegraphics[trim=0.cm 0.0cm 0.cm 0.0cm, clip=true, width=.8in]{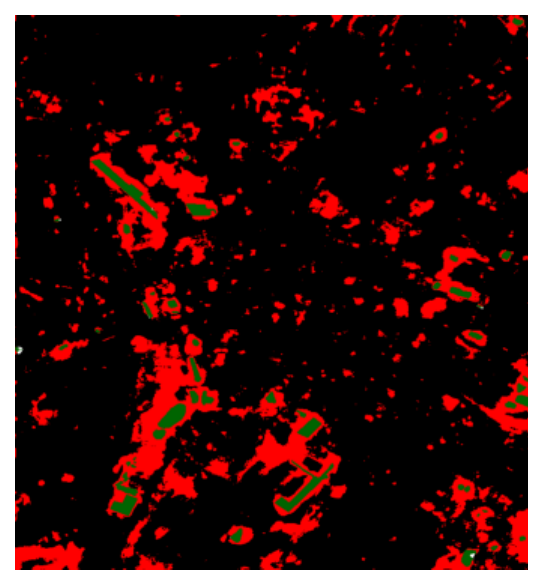} \\ 
\vspace*{-0pt}
\end{minipage}

\begin{minipage}{0.12\textwidth}
\includegraphics[trim=0.cm 0.0cm 0.cm 0cm, clip=true, width=0.77in]{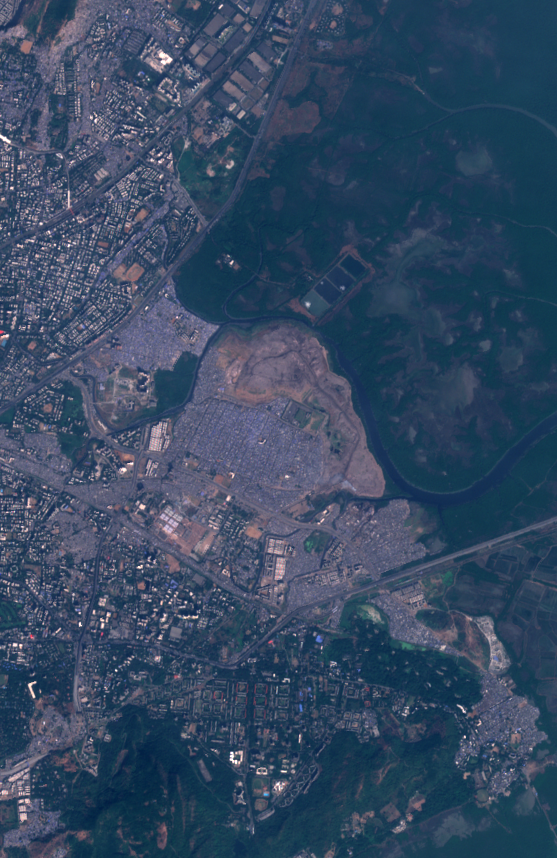} \\ 
\vspace*{-0pt}
\end{minipage}
\begin{minipage}{0.12\textwidth}
\includegraphics[trim=0cm 0cm 0cm 0cm, clip=true, width=0.8in]{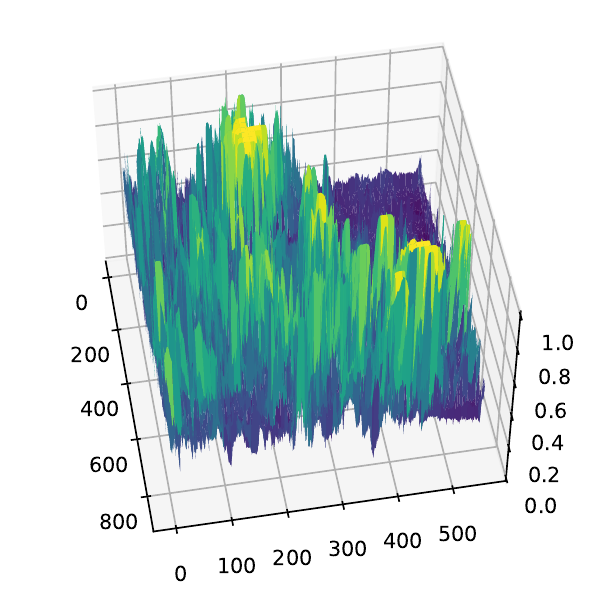} \\ 
\vspace*{-5pt}
\end{minipage}
\begin{minipage}{0.12\textwidth}
\includegraphics[trim=0.cm 0.0cm 0.cm 0cm, clip=true, width=0.77in]{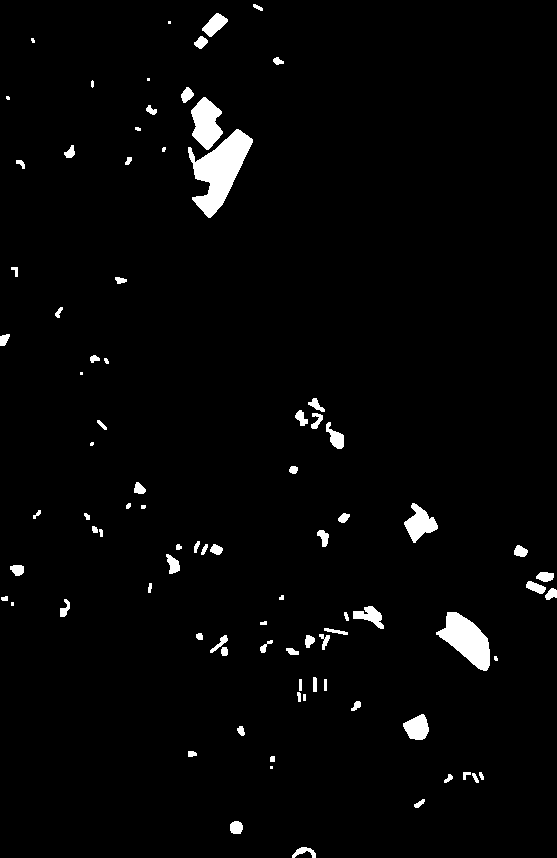} \\ 
\vspace*{-0pt}
\end{minipage}
\begin{minipage}{0.12\textwidth}
\includegraphics[trim=0.cm 0.0cm 0.cm 0cm, clip=true, width=0.8in]{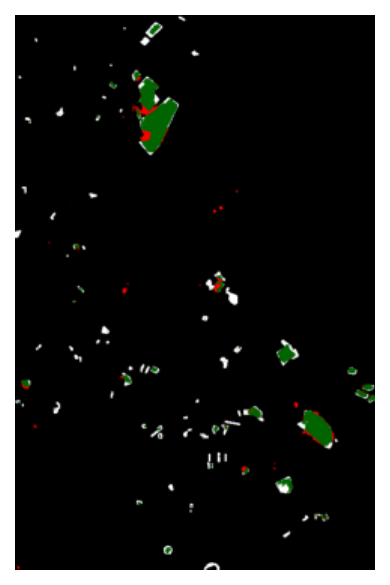} \\ 
\vspace*{-0pt}
\end{minipage}
\begin{minipage}{0.12\textwidth}
\includegraphics[trim=0.cm 0.0cm 0.cm 0.0cm, clip=true, width=.8in]{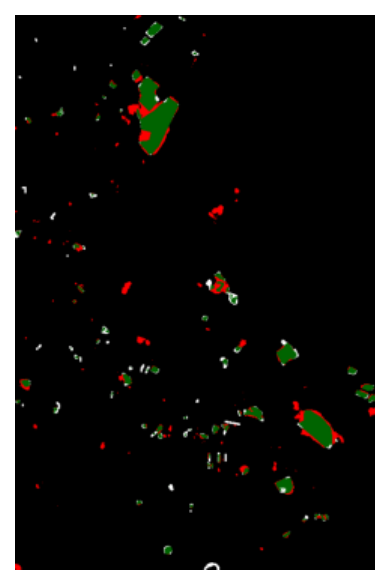} \\ 
\vspace*{-0pt}
\end{minipage}
\begin{minipage}{0.12\textwidth}
\includegraphics[trim=0.cm 0.0cm 0.cm 0.0cm, clip=true, width=.8in]{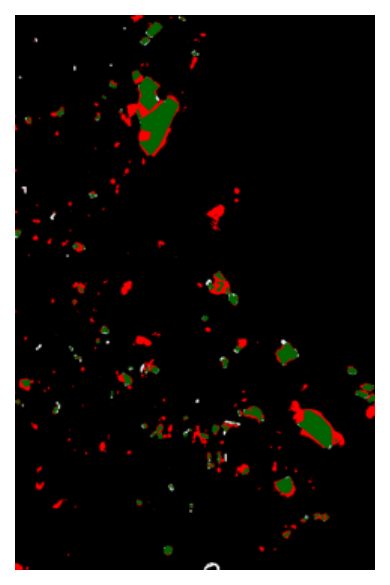} \\ 
\vspace*{-0pt}
\end{minipage}
\begin{minipage}{0.12\textwidth}
\includegraphics[trim=0.cm 0.0cm 0.cm 0.0cm, clip=true, width=.8in]{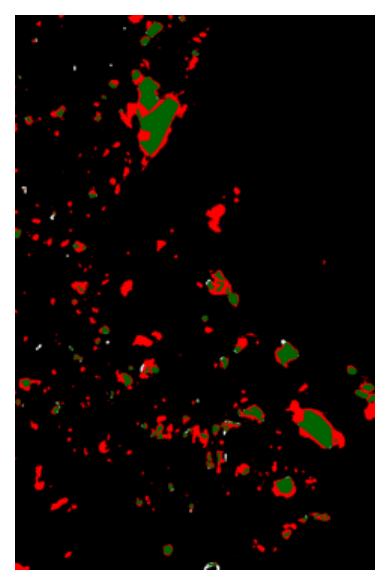} \\ 
\vspace*{-0pt}
\end{minipage}
\begin{minipage}{0.12\textwidth}
\includegraphics[trim=0.cm 0.0cm 0.cm 0.0cm, clip=true, width=.8in]{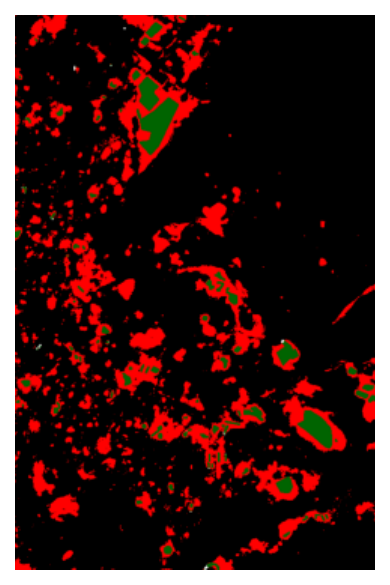} \\ 
\vspace*{-0pt}
\end{minipage}

\begin{minipage}{0.12\textwidth}
\includegraphics[trim=0.cm 0.0cm 0.cm 0cm, clip=true, width=0.77in]{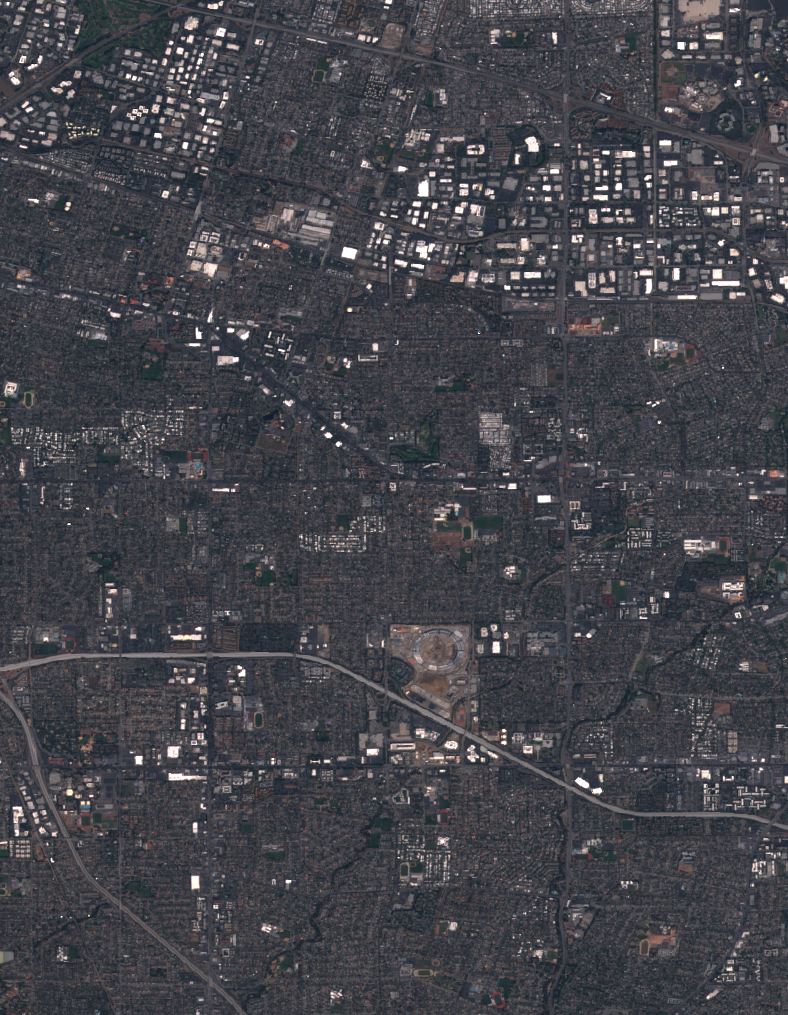} \\ 
\vspace*{-0pt}
\end{minipage}
\begin{minipage}{0.12\textwidth}
\includegraphics[trim=0cm 0cm 0cm 0cm, clip=true, width=0.8in]{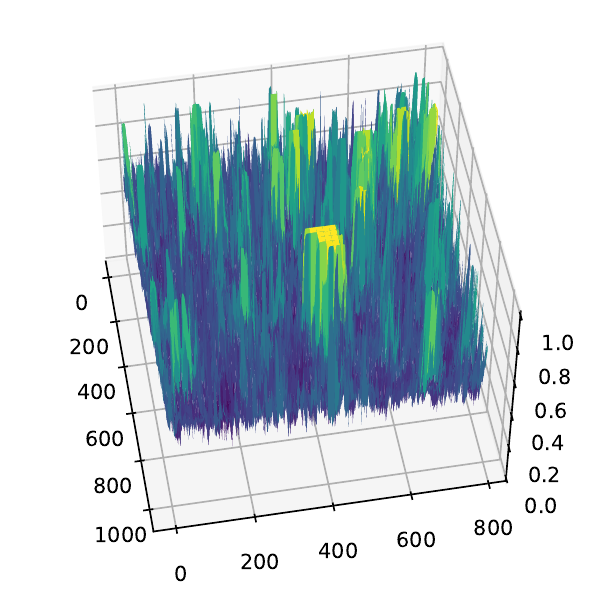} \\ 
\vspace*{-5pt}
\end{minipage}
\begin{minipage}{0.12\textwidth}
\includegraphics[trim=0.cm 0.0cm 0.cm 0cm, clip=true, width=0.77in]{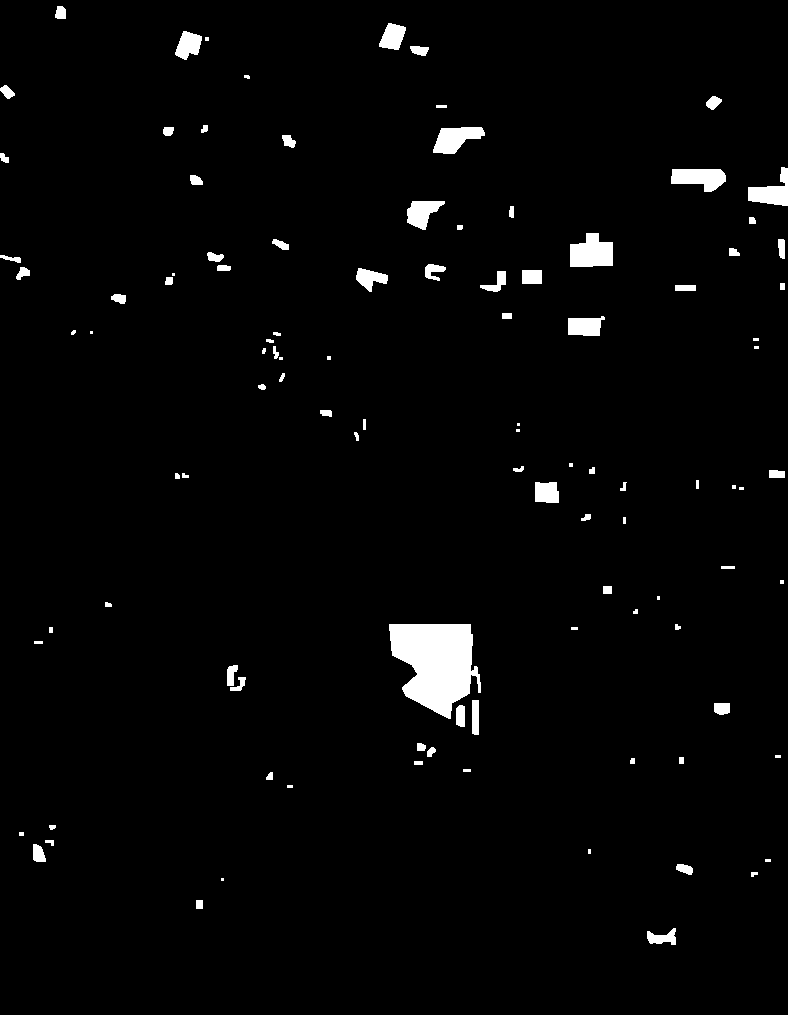} \\ 
\vspace*{-0pt}
\end{minipage}
\begin{minipage}{0.12\textwidth}
\includegraphics[trim=0.cm 0.0cm 0.cm 0cm, clip=true, width=0.8in]{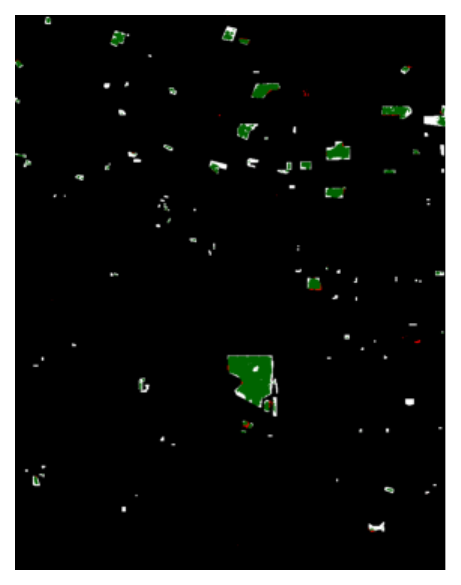} \\ 
\vspace*{-0pt}
\end{minipage}
\begin{minipage}{0.12\textwidth}
\includegraphics[trim=0.cm 0.0cm 0.cm 0.0cm, clip=true, width=.8in]{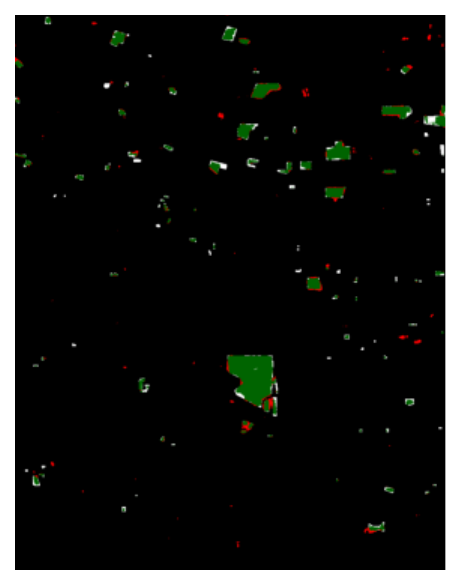} \\ 
\vspace*{-0pt}
\end{minipage}
\begin{minipage}{0.12\textwidth}
\includegraphics[trim=0.cm 0.0cm 0.cm 0.0cm, clip=true, width=.8in]{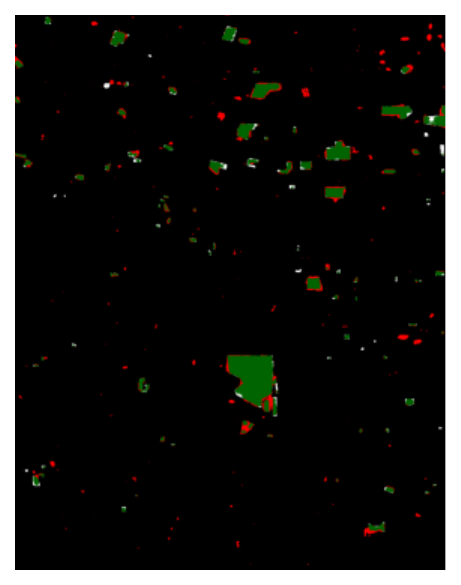} \\ 
\vspace*{-0pt}
\end{minipage}
\begin{minipage}{0.12\textwidth}
\includegraphics[trim=0.cm 0.0cm 0.cm 0.0cm, clip=true, width=.8in]{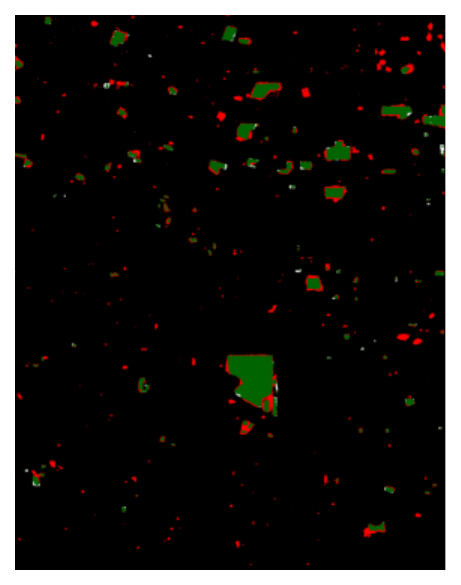} \\ 
\vspace*{-0pt}
\end{minipage}
\begin{minipage}{0.12\textwidth}
\includegraphics[trim=0.cm 0.0cm 0.cm 0.0cm, clip=true, width=.8in]{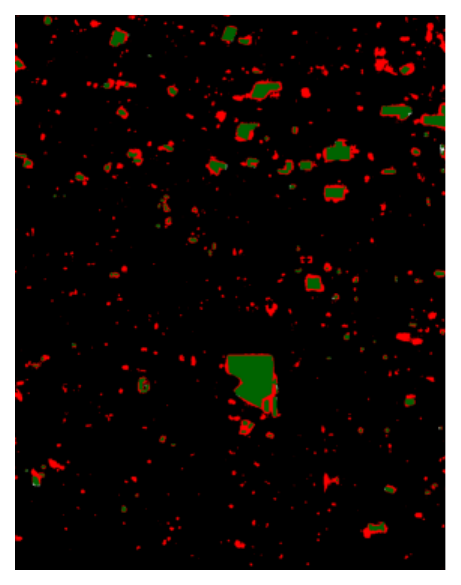} \\ 
\vspace*{-0pt}
\end{minipage}

\begin{minipage}{0.12\textwidth}
\includegraphics[trim=0.cm 0.0cm 0.cm 0cm, clip=true, width=0.77in]{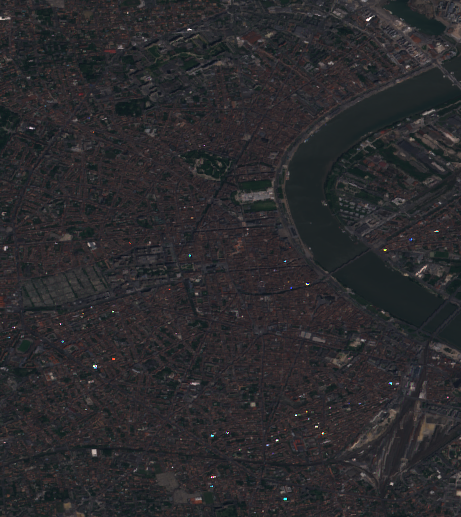} \\ 
\vspace*{-0pt}
\end{minipage}
\begin{minipage}{0.12\textwidth}
\includegraphics[trim=0cm 0cm 0cm 0cm, clip=true, width=0.8in]{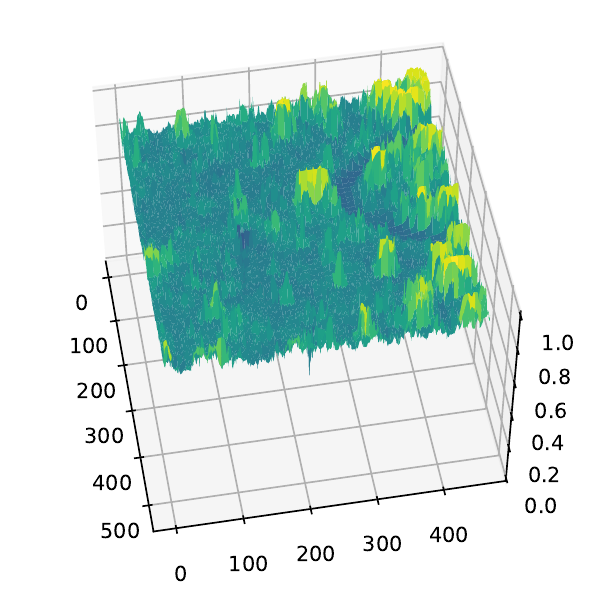} \\ 
\vspace*{-5pt}
\end{minipage}
\begin{minipage}{0.12\textwidth}
\includegraphics[trim=0.cm 0.0cm 0.cm 0cm, clip=true, width=0.77in]{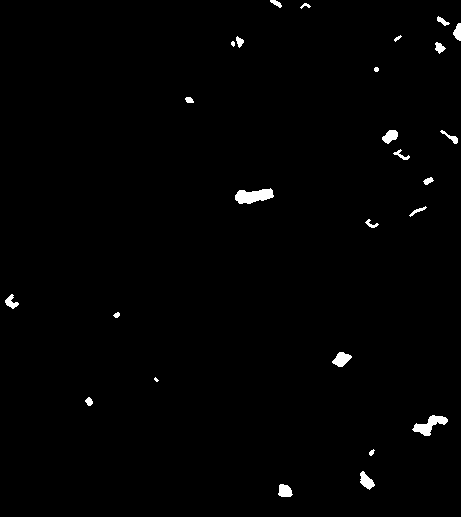} \\ 
\vspace*{-0pt}
\end{minipage}
\begin{minipage}{0.12\textwidth}
\includegraphics[trim=0.cm 0.0cm 0.cm 0cm, clip=true, width=0.8in]{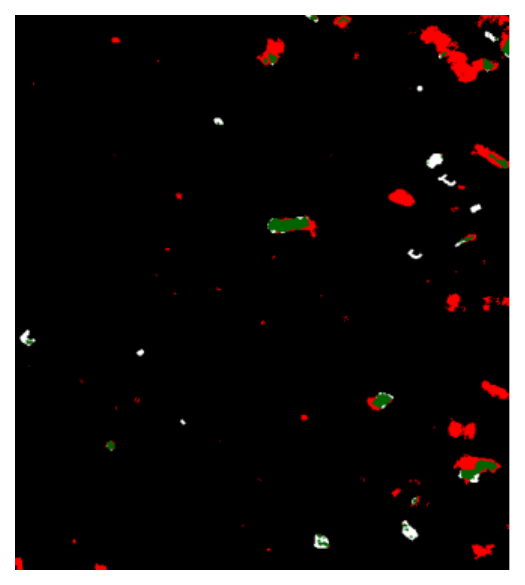} \\ 
\vspace*{-0pt}
\end{minipage}
\begin{minipage}{0.12\textwidth}
\includegraphics[trim=0.cm 0.0cm 0.cm 0.0cm, clip=true, width=.8in]{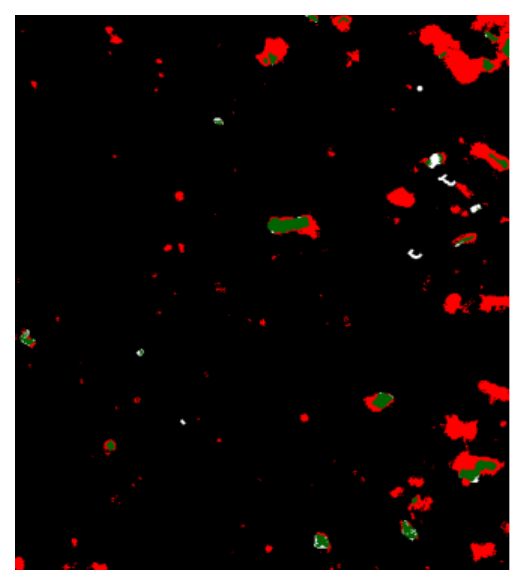} \\ 
\vspace*{-0pt}
\end{minipage}
\begin{minipage}{0.12\textwidth}
\includegraphics[trim=0.cm 0.0cm 0.cm 0.0cm, clip=true, width=.8in]{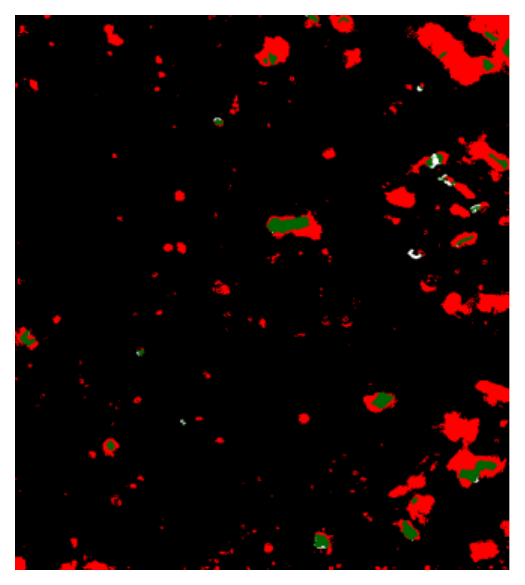} \\ 
\vspace*{-0pt}
\end{minipage}
\begin{minipage}{0.12\textwidth}
\includegraphics[trim=0.cm 0.0cm 0.cm 0.0cm, clip=true, width=.8in]{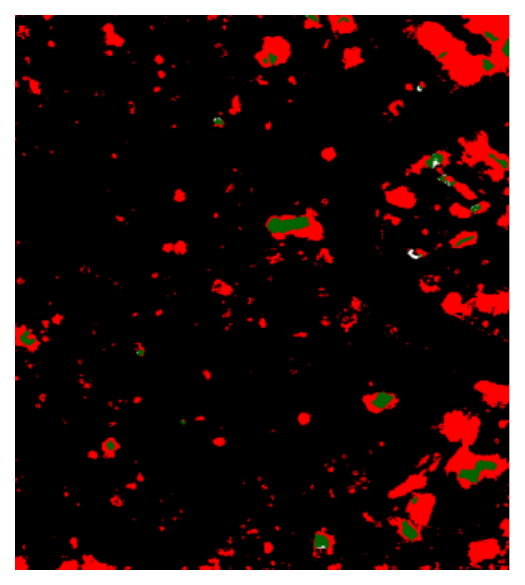} \\ 
\vspace*{-0pt}
\end{minipage}
\begin{minipage}{0.12\textwidth}
\includegraphics[trim=0.cm 0.0cm 0.cm 0.0cm, clip=true, width=.8in]{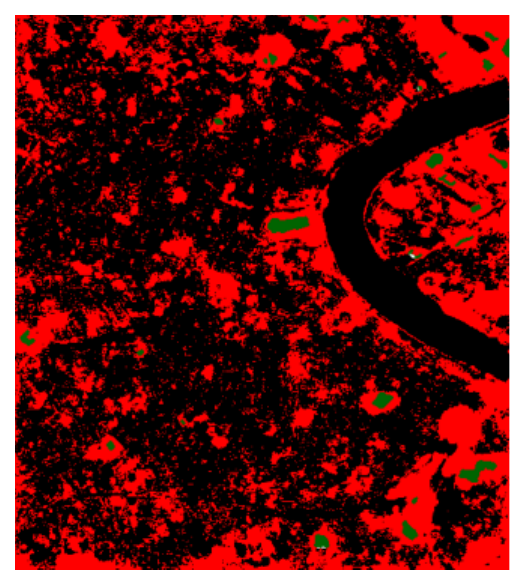} \\ 
\vspace*{-0pt}
\end{minipage}

\begin{minipage}{0.12\textwidth}
\includegraphics[trim=0.cm 0.0cm 0.cm 0cm, clip=true, width=0.77in]{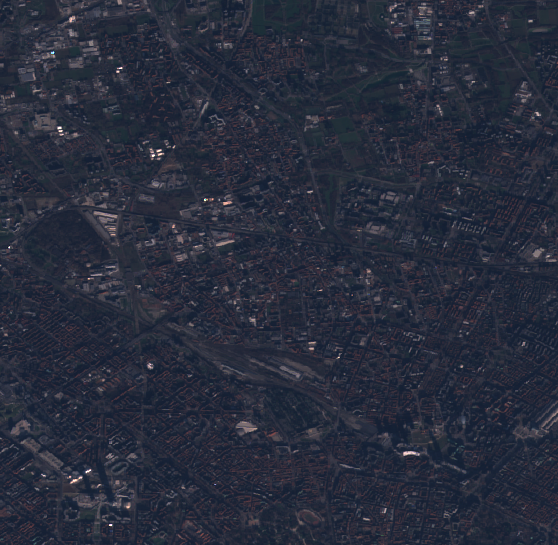} \\ 
\vspace*{-0pt}
\centering{$(a)$ }
\end{minipage}
\begin{minipage}{0.12\textwidth}
\includegraphics[trim=0cm 0cm 0cm 0cm, clip=true, width=0.8in]{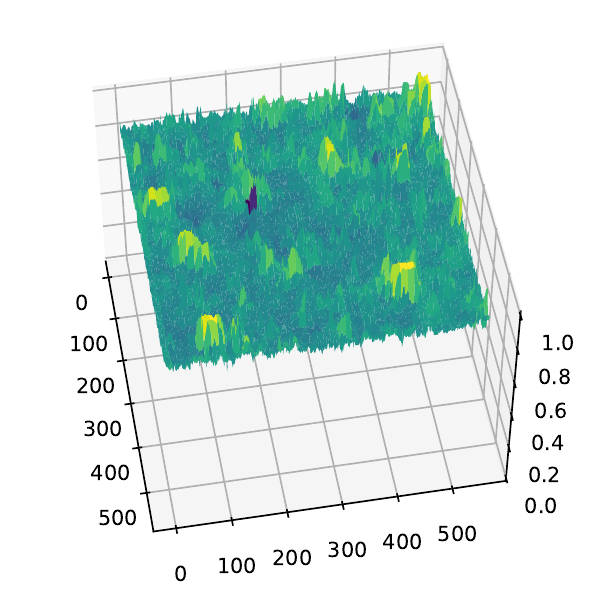} \\ 
\vspace*{-5pt}
\centering {$(b)$ }
\end{minipage}
\begin{minipage}{0.12\textwidth}
\includegraphics[trim=0.cm 0.0cm 0.cm 0cm, clip=true, width=0.77in]{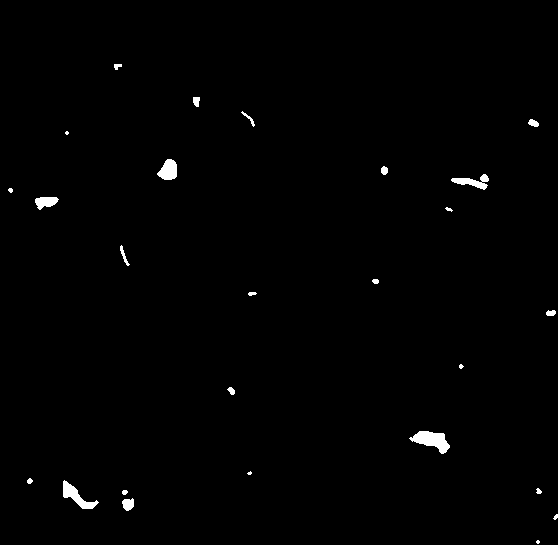} \\ 
\vspace*{-0pt}
\centering{$(c)$}
\end{minipage}
\begin{minipage}{0.12\textwidth}
\includegraphics[trim=0.cm 0.0cm 0.cm 0cm, clip=true, width=0.8in]{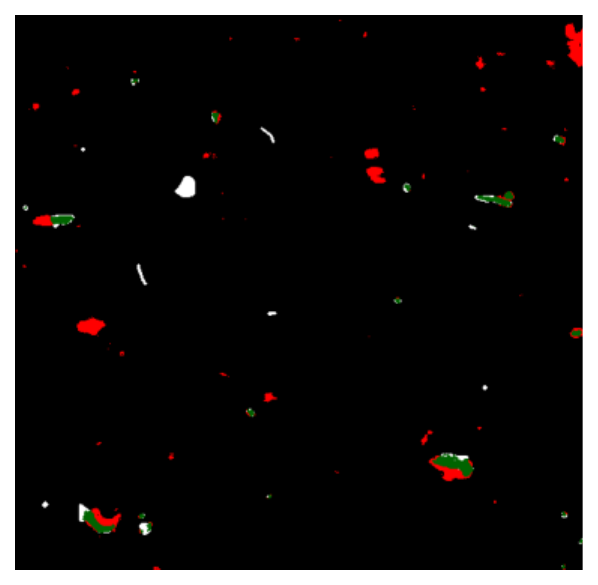} \\ 
\vspace*{-0pt}
\centering {$(d)$}
\end{minipage}
\begin{minipage}{0.12\textwidth}
\includegraphics[trim=0.cm 0.0cm 0.cm 0.0cm, clip=true, width=.8in]{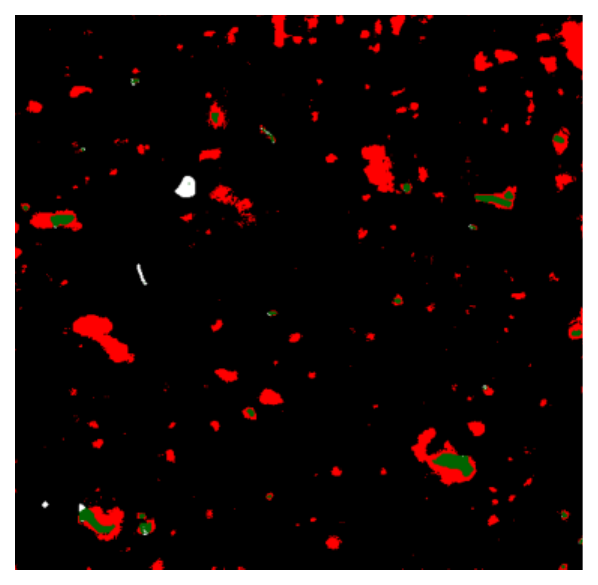} \\ 
\vspace*{-0pt}
\centering {$(e)$ }
\end{minipage}
\begin{minipage}{0.12\textwidth}
\includegraphics[trim=0.cm 0.0cm 0.cm 0.0cm, clip=true, width=.8in]{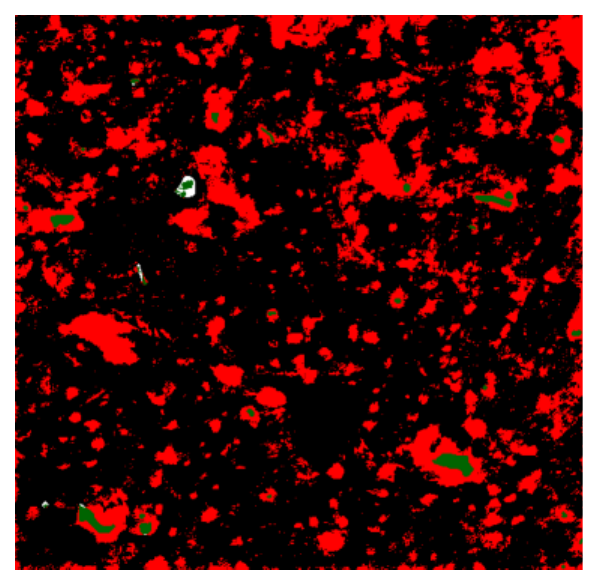} \\ 
\vspace*{-0pt}
\centering {$(f)$ }
\end{minipage}
\begin{minipage}{0.12\textwidth}
\includegraphics[trim=0.cm 0.0cm 0.cm 0.0cm, clip=true, width=.8in]{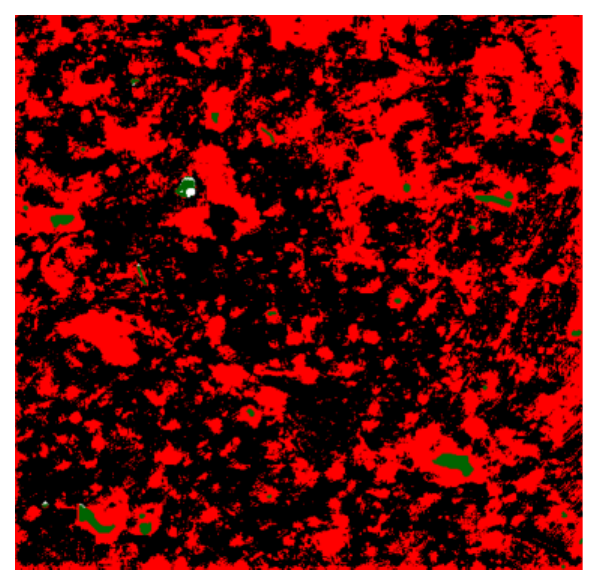} \\ 
\vspace*{-0pt}
\centering {$(g)$ }
\end{minipage}
\begin{minipage}{0.12\textwidth}
\includegraphics[trim=0.cm 0.0cm 0.cm 0.0cm, clip=true, width=.8in]{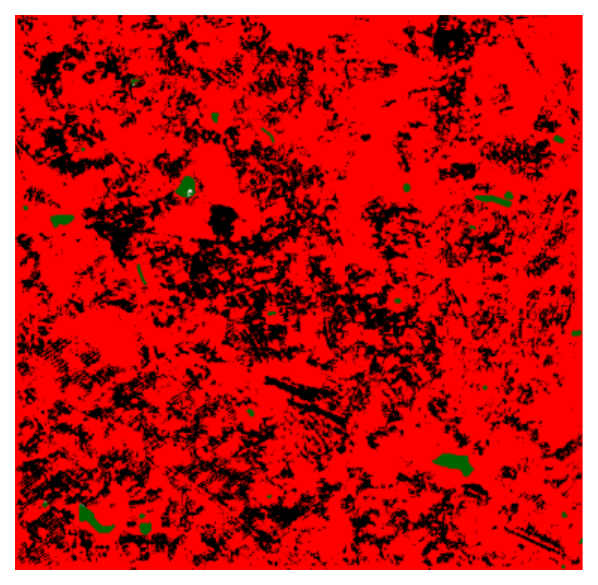} \\ 
\vspace*{-0pt}
\centering {$(h)$ }
\end{minipage}

\caption{Visualization of reference and change detection maps across change encoding rates. (a) The reference image displayed in RGB format.
(b) The scoring map representing the intensity of detected changes.
(c) The ground truth change map for validation.
(d-h) Encoding maps showing change detection results with varying change encoding rate of 60\%, 80\%, 90\%, 95\%, and 99\%, respectively.
In the encoding maps, red pixels represent encoded changes, green pixels represent true positives, and white pixels represent false negatives.}
\label{fig_change}
\vspace*{-12pt}
\end{figure*}
Fig.~\ref{fig_change} illustrates the change detection results in varying change encoding rates, showcasing the adaptability and limitations of the proposed semantic encoding framework. Fig.~\ref{fig_change}(a) shows the reference image in RGB format, serving as the baseline for detecting changes, while Fig.~\ref{fig_change}(b) displays the scoring map, where intensity indicates the likelihood of changes, enabling prioritization of high-probability regions. Fig.~\ref{fig_change}(c) provides the ground truth change map for validation. Figs.~\ref{fig_change}(d)–(h) illustrate encoding maps at rates of 60\%, 80\%, 90\%, 95\%, and 99\%, respectively, highlighting the trade-off between encoding efficiency and detection quality. Interestingly, in the first four cases, the probability map effectively isolates changed information, enabling a more selective approach that minimizes redundant transmission. However, in other cases, the algorithm prioritizes transmitting more overlapping information to preserve the quality of the changed pixels selected, particularly when lower encoding rates are used. This shows the ability of the algorithm to balance between preserving critical data and achieving higher compression, although it also underscores the challenge of optimizing trade-offs for different transmission constraints. 

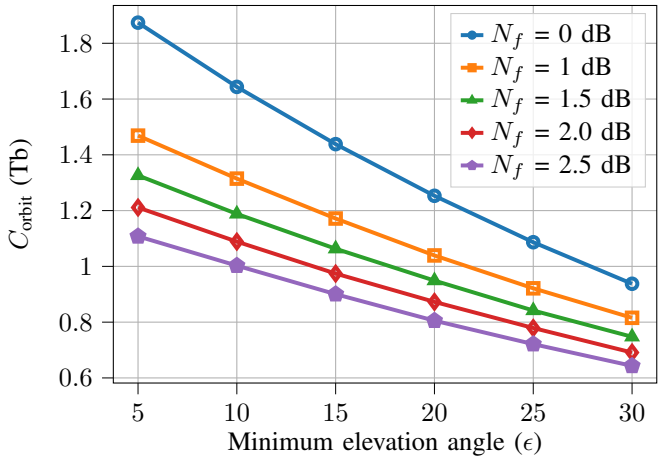
\begin{figure}[!t]
    \centering
\begin{tikzpicture}

\definecolor{crimson2143940}{RGB}{214,39,40}
\definecolor{darkgray176}{RGB}{176,176,176}
\definecolor{darkorange25512714}{RGB}{255,127,14}
\definecolor{forestgreen4416044}{RGB}{44,160,44}
\definecolor{lightgray204}{RGB}{204,204,204}
\definecolor{mediumpurple148103189}{RGB}{148,103,189}
\definecolor{steelblue31119180}{RGB}{31,119,180}

\begin{axis}[
  height=0.75\linewidth, 
  width=1.0\linewidth,
  legend image post style={scale=0.5}, 
legend cell align={left},
legend style={fill opacity=0.8, draw opacity=1, text opacity=1, draw=lightgray204},
tick align=outside,
tick pos=left,
x grid style={darkgray176},
xlabel={Minimum elevation angle (\(\displaystyle \epsilon\))},
xmajorgrids,
xmin=3.75, xmax=31.25,
xtick style={color=black},
y grid style={darkgray176},
ylabel={$C_\text{orbit}$ (Tb)},
ymajorgrids,
ymin=0.581716371038509, ymax=1.93550212536787,
ytick style={color=black}
]
\addplot [semithick, steelblue31119180, solid, mark=o, mark size=2pt, mark options={solid}, line width=1.5pt]
table {%
5 1.87396640926199
10 1.64361180714356
15 1.4383081304521
20 1.25307624225994
25 1.08661342146031
30 0.937584392332171
};
\addlegendentry{$N_f$ = 0 dB}
\addplot [semithick, darkorange25512714, solid, mark=square, mark size=2pt, mark options={solid}, line width=1.5pt]
table {%
5 1.46881662625157
10 1.31465211566065
15 1.1715709272553
20 1.03919981383766
25 0.921301043538869
30 0.815543943636279
};
\addlegendentry{$N_f$ = 1 dB}
\addplot [semithick, forestgreen4416044,  solid, mark=triangle*, mark size=2pt, mark options={solid}, line width=1.5pt]
table {%
5 1.32610633794819
10 1.18809703165276
15 1.06320931614537
20 0.948976622577323
25 0.841695377934588
30 0.747466906766684
};
\addlegendentry{$N_f$ = 1.5 dB}
\addplot [semithick, crimson2143940,  solid, mark=diamond, mark size=2.5pt, mark options={solid}, line width=1.5pt]
table {%
5 1.21103139687557
10 1.08876539634853
15 0.974184900020281
20 0.872659876632033
25 0.779075059150357
30 0.691078358139095
};
\addlegendentry{$N_f$ = 2.0 dB}
\addplot [semithick, mediumpurple148103189,  solid, mark = pentagon*, mark size=2.5pt, mark options={solid}, line width=1.5pt]
table {%
5 1.1076942090715
10 1.00261102027388
15 0.900353090006204
20 0.805284568541919
25 0.721238049334234
30 0.643252087144389
};
\addlegendentry{$N_f$ = 2.5 dB}
\end{axis}

\end{tikzpicture}
    \vspace{-10pt}
    \caption{Achievable data rates for satellite-to-ground communication in one orbit.}
    \label{fig:capacity}
    \vspace{-10pt}
\end{figure}
Fig.~\ref{fig:capacity} illustrates the achievable data rates ($C_{\text{orbit}}$) for satellite-to-ground communication over one orbit as a function of the minimum elevation angle ($\epsilon$). The results demonstrate a clear inverse relationship between $\epsilon$ and $C_{\text{orbit}}$, with data rates decreasing significantly as the elevation angle increases from $5^\circ$ to $30^\circ$. At $\epsilon = 5^\circ$, the maximum $C_{\text{orbit}}$ reaches approximately 1.8 Tb for $N_f = 0$ dB, whereas it drops to around 0.9 Tb at $\epsilon = 30^\circ$ dB. Similarly, for higher $N_f$, such as $N_f = 2.5$ dB, the reduction in $C_{\text{orbit}}$ is even more pronounced across all elevation angles. These results give us information to analyze and predict the orbit capacity effectively.

\begin{figure}[!t]
    \centering
\begin{tikzpicture}

\definecolor{darkblue}{RGB}{0,0,139}
\definecolor{darkgray176}{RGB}{176,176,176}
\definecolor{darkorange}{RGB}{255,140,0}
\definecolor{lightgray204}{RGB}{204,204,204}

\begin{axis}[
height=0.7\linewidth, 
width=0.86\linewidth,
tick align=outside,
tick pos=left,
x grid style={darkgray176},
xlabel={Change encoding rate},
xmajorgrids,
xmin=0.564, xmax=1.026,
xtick={0.6, 0.7, 0.8, 0.9, 0.95, 0.99},
xticklabel style={anchor=north, rotate=30},
xtick style={color=black},
y grid style={darkgray176},
ylabel=\textcolor{blue}{Amount encoded data (Tb)},
ymin=0, ymax=0.749783484123742,
ytick style={color=black}
]
\draw[draw=none,fill=darkblue,fill opacity=0.9,postaction={pattern=sixpointed stars}] (axis cs:0.585,0) rectangle (axis cs:0.615,0.0588044014504544);
\draw[draw=none,fill=darkblue,fill opacity=0.9,postaction={pattern=sixpointed stars}] (axis cs:0.685,0) rectangle (axis cs:0.715,0.0906664917546886);
\draw[draw=none,fill=darkblue,fill opacity=0.9,postaction={pattern=sixpointed stars}] (axis cs:0.785,0) rectangle (axis cs:0.815,0.14548282681609);
\draw[draw=none,fill=darkblue,fill opacity=0.9,postaction={pattern=sixpointed stars}] (axis cs:0.885,0) rectangle (axis cs:0.915,0.304670871448037);
\draw[draw=none,fill=darkblue,fill opacity=0.9,postaction={pattern=sixpointed stars}] (axis cs:0.935,0) rectangle (axis cs:0.965,0.489129648078969);
\draw[draw=none,fill=darkblue,fill opacity=0.9,postaction={pattern=sixpointed stars}] (axis cs:0.975,0) rectangle (axis cs:1.005,0.714079508689278);
\end{axis}

\begin{axis}[
height=0.7\linewidth, 
width=0.86\linewidth,
axis y line=right,
legend cell align={left},
legend style={
  fill opacity=0.8,
  draw opacity=1,
  text opacity=1,
  at={(0.03,0.97)},
  anchor=north west,
  draw=lightgray204,
  axis line style={-}
},
tick align=outside,
x grid style={darkgray176},
xmin=0.564, xmax=1.026,
xtick={0.6, 0.7, 0.8, 0.9, 0.95, 0.99},
xticklabel style={anchor=north, rotate=30},
xtick pos=left,
xtick style={color=black},
y grid style={darkgray176},
ylabel=\textcolor{darkorange}{Average PSNR (dB)},
ymin=21.8706948216904, ymax=34.3455951537189,
ytick={22, 24, 26, 28, 30, 32, 34},
ytick pos=right,
ytick style={color=black},
yticklabel style={anchor=west}
]
\addplot [semithick, darkorange, mark=*, mark size=2, mark options={solid}, line width=1.5pt]
table {%
0.6 23.8790029559296
0.7 24.26655330228
0.8 24.859749295581
0.9 26.6109421660678
0.95 28.8768565162054
0.99 33.7785542295358
};
\addlegendentry{Proposed}
\addplot [semithick, darkorange, dashed, mark=square*, mark size=2, mark options={solid}, line width=1.5pt]
table {%
0.6 22.4377357458736
0.7 22.6111454210411
0.8 22.9136048681349
0.9 24.1206278171674
0.95 26.1235022052574
0.99 30.6342457831747
};
\addlegendentry{Baseline}
\end{axis}

\end{tikzpicture}
    \vspace{-10pt}
    \caption{Impact of change encoding rate on data volume and quality.}
    \label{fig5_final}
    \vspace{-10pt}
\end{figure}
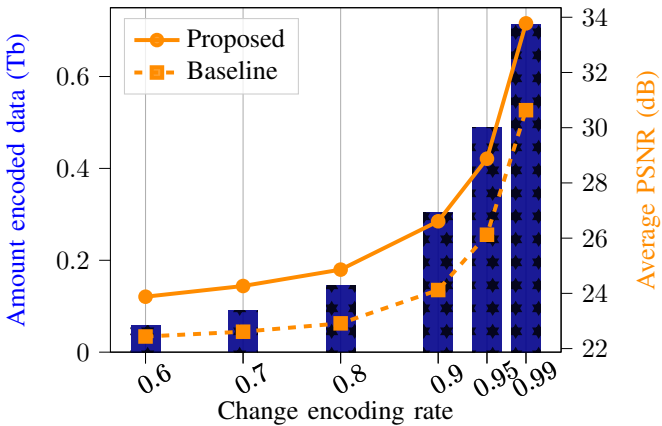
Finally, Fig.~\ref{fig5_final} shows the relationship between change encoding rate, encoded data, and PSNR for the proposed approach compared to a baseline that randomly selects pixels using the same selected data volume. The proposed method achieves a steep PSNR increase from 24 dB at 0.6 accuracy to 34 dB at 0.99, while the baseline consistently lags by around 2 dB. For instance, at 0.8 accuracy, the proposed approach reaches 22.9 dB, compared to 24.86 dB for the baseline. The results highlight the trade-off between data quality and the volume of encoded data in satellite communication, illustrating the sacrifices in data quality required to reduce transmission volume. Moreover, they demonstrate that the proposed method effectively prioritizes critical pixels, achieving superior data quality compared to the baseline while maintaining the same transmission volume.

\section{Conclusion}
This paper proposed an adaptive framework for efficient downlink communication of multi-spectral satellite images, leveraging deep learning-based semantic encoding and channel capacity prediction.  Unlike traditional approaches, the framework prioritizes critical information and dynamically adjusts to fluctuating communication conditions, addressing limitations of channel capacities. Experimental results demonstrated high-quality data delivery with enhanced resource efficiency, making it a promising solution for EO applications. 
\setstretch{0.95}
\bibliographystyle{IEEEtran}
\bibliography{journal}
\end{document}